\documentclass{ws-procs975x65}
\usepackage{graphicx}
\newcommand{\be}{\begin{equation}}
\newcommand{\ee}{\end{equation}}
\newcommand{\een}{\end{subequations}}
\newcommand{\ben}{\begin{subequations}}
\newcommand{\beq}{\begin{eqalignno}}
\newcommand{\eeq}{\end{eqalignno}}

\newcommand{\lsim}{\mathrel{\mathop{\kern 0pt \rlap
      {\raise.2ex\hbox{$<$}}}\lower.9ex\hbox{\kern-.190em $ \sim$}}}
\newcommand{\gsim}{\mathrel{\mathop{\kern 0pt
      \rlap{\raise.2ex\hbox{$>$}}}\lower.9ex\hbox{\kern-.190em $\sim$}}}

\def\beq{\begin{equation}}
\def\eeq{\end{equation}}
\begin{document}

\title{New approaches in the analysis of Dark Matter direct detection
data: scratching below the surface of the most general WIMP parameter
space.}
\author{Scopel, Stefano$^*$; Yoon, Kook-Hyun; Yoon, Jong-Hyun}

\address{Physics Department, Sogang University,\\
Seoul, 121-742, South Korea\\
$^*$E-mail: scopel@sogang.ac.kr}

\begin{abstract}
  We show that compatibility between the DAMA modulation result (as
  well as less statistically significant excesses such as the CDMS
  Silicon effect and the excess claimed by CRESST) with constraints
  from other experiments can be achieved by extending the analysis of
  direct detection data beyond the standard elastic scattering of a
  WIMP off nuclei with a spin--dependent or a spin--independent cross
  section and with a velocity distribution as predicted by the
  Isothermal Sphere model. To do so we discuss several new approaches
  for the analysis of Dark Matter direct detection data, with the goal
  to remove or reduce its dependence on specific theoretical
  assumptions, and to extend its scope: the factorization approach of
  astrophysics uncertainties, the classification and study of
  WIMP--nucleon interactions within non--relativistic field theory,
  inelastic scattering and isovector-coupling cancellations including
  subdominant two--nucleon NLO effects. Typically, combining two or
  more of these ingredients can lead to conclusions which are very
  different to what usually claimed in the literature. This shows that
  we are only starting now to scratch the surface of the most general
  WIMP direct detection parameter space.
\end{abstract}

\keywords{Dark Matter; direct detection}

\bodymatter

%%%%%%%%%%%%%%%%% now a standard article style for the most part

\section{Content}

Many underground experiments are currently searching for Weakly
Interacting Massive Particles (WIMPs), which are the most popular
candidates to provide the Dark Matter (DM) which is believed to make
up 27\% of the total mass density of the Universe, and more than 90\%
of the halo of our Galaxy.  One of them (DAMA\cite{dama}) has been
observing for more than 15 years a yearly modulation effect in the low
part of its energy spectrum which is consistent with that expected due
to the Earth rotation around the Sun from the elastic scattering of
WIMPs off the sodium iodide nuclei that constitute the crystals of its
scintillators.  Many experimental collaborations using nuclear targets
different from $NaI$ and various background--subtraction techniques to
look for WIMP--elastic scattering (including LUX\cite{lux},
SuperCDMS\cite{super_cdms}, COUPP\cite{coupp}, PICASSO\cite{picasso})
have failed to observe any anomaly so far, implying severe constraints
on the most popular WIMP scenarios used to explain the DAMA excess.  A
similar situation arises when confronting the excesses claimed by
CDMS-Si \cite{cdms_si} ad CRESST\cite{cresst} with the same
constraints.

However, besides the fact that several experimental uncertainties
might still be advocated to question the robustness of these bounds,
in most cases such conclusions are drawn by analysing direct detection
data assuming elastic scattering of a WIMP off nuclei with a
spin--dependent or a spin--independent cross section and with a WIMP
velocity distribution $f(\vec{v})$ as predicted by the Isothermal
Sphere model.  These assumtions may be well motivated, but represent
only a very small part of the possible and experimentally viable
options.

For instance, if the DAMA effect can be explained by scatterings of
WIMPS off sodium such constraints can be potentially relaxed by
considering models where the nuclear response function of sodium on
the WIMP interaction is enhanced compared to that of the nuclei used
to obtain the experimental bounds (germanium, xenon, fluorine).

In light of the situation summarized above several new directions have
been explored in the recent past both to remove as much as possible
the dependence on specific theoretical assumptions (both of
particle--physics and astrophysical origin) from the analysis of DM
direct detection data and to extend its scope to a wider class of
models. 

Starting from \cite{factorization} a general strategy has been
developed to factor out the dependence on $f(\vec{v})$ of the expected
WIMP--nucleus differential rate $dR/d E_R$ at the given recoil energy
$E_R$. This approach exploits the fact that $dR/d E_R$ depends on
$f(\vec{v})$ only through the minimal velocity $v_{min}$ that the WIMP
must have to deposit at least $E_R$, i.e.:
\begin{equation}
\frac{dR}{d E_R}\propto \eta(v_{min})\equiv \int_{|\vec{v}|>v_{min}}\frac{f(\vec{v})}{|\vec{v}|}\; d^3 v.
\label{eq:eta_tilde_ex}
\end{equation}
\noindent By mapping recoil energies $E_R$ into same ranges of
$v_{min}$ the dependence on $\eta(v_{min})$ and so on $f(\vec{v})$
cancels out in the ratio of expected rates on different targets.
Since the mapping between $E_R$ and $v_{min}$ depends on the nuclear
mass the factorization of $\eta(v_{min})$ is only possible in the case
of detectors with a single nuclear target, or for which the expected
rate is dominated by scatterings on a single target. In
\cite{noi1,noi3}, after discussing the applicability and limitations
of this method to the DAMA data, we have extended this procedure to
the case when a constraining experiment contains different targets.

\begin{figure}
\begin{center}
\includegraphics[width=0.49\columnwidth,bb= 46 194 506 635]{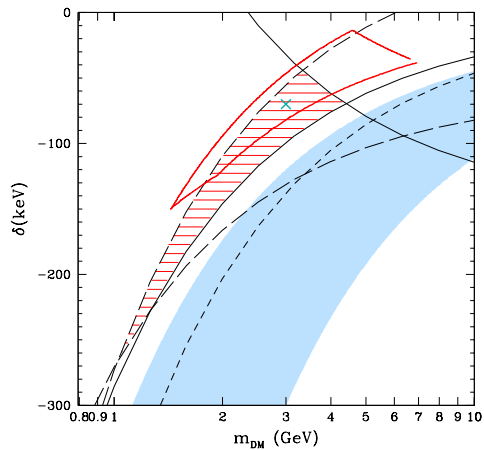}
\end{center}
\caption{Mass splitting $\delta=m^{\prime}_{DM}-m_{DM}$ as a function
  of $m_{DM}$.  The closed contour bounded by the solid (red) line
  represents the IDM parameter space where the modulation effect
  measured by DAMA is compatible to other constraints, while the
  horizontally (red) hatched area represents the same for the excess
  measured by CDMS-$Si$ (see Ref.\cite{noi1} for details). In both
  cases $f_n/f_p$=-0.79.}
\label{fig:halo_independent}
\end{figure}

In particular, a scenario proposed to alleviate the tension among
different direct detection experiments is Inelastic Dark Matter (IDM).
In this class of models a Dark Matter (DM) particle $\chi$ of mass
$m_{DM}$ interacts with atomic nuclei exclusively by up--scattering to
a second state $\chi^{\prime}$ with mass
$m_{DM}^{\prime}=m_{DM}+\delta$.  In the case of exothermic Dark
Matter $\delta<0$ is also possible: in this case the particle $\chi$
is metastable and down--scatters to a lighter state
$\chi^{\prime}$. In the IDM scenario the halo--model factorization
approach is more complicated that in the elastic case, because in
presence of a mass splitting $\delta\ne$0 the mapping between the
nuclear recoil energy $E_R$ and the minimal velocity $v_{min}$ that
the incoming WIMP needs to have to deposit $E_R$ becomes more involved
than in the elastic case.  In \cite{noi1} we provided the first
systematic analysis of IDM for a spin--independent cross section, i.e. scaling as:
\begin{equation}
\sigma\propto \left [ Z f_p+(A-Z) f_n \right ]^2,
\label{eq:spin_independent}
\end{equation}
\noindent with $f_{p,n}$ the WIMP couplings to protons and neutrons,
respectively and $A,Z$ the nuclear mass and atomic number.  Including
all available data and making use of the factorization property of the
halo--model dependence we introduced some strategies to determine
regions in the IDM parameter space where the tension existing among
different experimental results can be eliminated or at least
alleviated: indeed, as shown in Fig.\ref{fig:halo_independent},
compatibility between DAMA or CDMS-Si and all other constraints can be
achieved. In the same analysis, we also showed that the same thing can
be also obtained for the excess measured by CRESST \cite{cresst} (we
note there that this result, which holds for $m_{\chi}\gsim$ 40 GeV,
is still true after the low--background re-analysis of
Ref.\cite{cresst2}).
 
%%%===========================================
%%%===========================================

%%%===========================================
\begin{figure}
\begin{center}
\includegraphics[width=0.49\columnwidth,bb= 46 194 506 635]{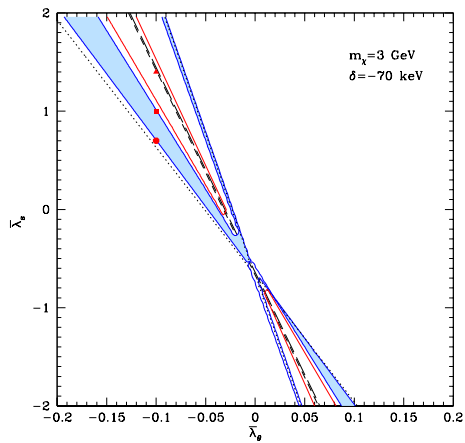}
\end{center}
\caption{In this plot for any choice of $\bar{\lambda}_{\theta}$ and
  $\bar{\lambda}_s$, which parametrize the metastable WIMP couplings
  to heavy quarks and the strange quark, respectively, the ratio of
  the corresponding couplings to light quarks
  $\bar{\lambda}_d/\bar{\lambda}_u$ is fixed to maximize the
  compatibility between CDMS--Si and SuperCDMS.  The shaded regions
  represent the parameter space where CDMS--Si and SuperCDMS are
  mutually compatible including NLO corrections to the cross section
  while at the same time the metastable state $\chi$ can be a thermal
  relic (i.e. $\Omega_{\chi}h^2\le$0.12).  The dotted curve represents
  the condition $\Omega_{\chi}h^2$=0.12 calculated using the LO
  instead of the NLO scaling law for the expected rate. The inner
  solid (red) line corresponds to $\tau=1/\Gamma$=4$\times$10$^{26}$
  seconds, the lifetime of the metastable state $\chi$. See
  Ref.\cite{noi2} for details.}
\label{fig:isospin_violation}
\end{figure}
%%%===========================================

On the other hand, in \cite{noi2} we have considered the effect of the
inclusion of the NLO corrections calculated in
\cite{isospin_violation_nlo} in a specific scenario of light IDM with
the same scaling of the cross section as in
Eq. (\ref{eq:spin_independent}) for which $f_n,f_p$ violate isospin
symmetry (Isospin--violating Dark Matter, IVDM) leading to a
suppression of the WIMP cross section off Germanium targets. By
incorporating this scenario in the halo--independent approach also in
this case, as shown in Fig.\ref{fig:isospin_violation}, a region of
the parameter space can be achieved where the CDMS-Si excess is
compatible to other constraint. This is obtained for $m_{\chi}\lsim4$
GeV and $\delta<0$ (exothermic DM).

One of the most popular scenarios for WIMP--nucleus scattering is a
spin--dependent interaction where the WIMP particle $\chi$ is a
fermion (either Dirac or Majorana) that recoils on the target nucleus
$T$ through it coupling to the spin $\vec{S}_N$ of nucleons $N=(p,n)$:
\begin{equation} {\cal L}_{int}\propto \vec{S}_{\chi}\cdot
  \vec{S}_N=c^p\vec{S}_{\chi}\cdot \vec{S}_p+c^n\vec{S}_{\chi}\cdot
  \vec{S}_n.
\label{eq:spin_dependent}
\end{equation}
Among the main motivations of such scenario is the fact that the most
stringent bounds on the interpretation of the DAMA effect in terms of
WIMP--nuclei scatterings arise today from detectors using xenon
(LUX\cite{lux}) and germanium (SuperCDMS\cite{super_cdms}) whose spin
is mostly originated by an unpaired neutron, while both sodium and
iodine in DAMA have an unpaired proton: if the WIMP effective coupling
to neutrons $c^n$ is suppressed compared to that on protons $c^p$ this
class of bounds can be evaded. However this scenario is presently
constrained by droplet detectors (including COUPP\cite{coupp}) and
bubble chambers (including PICASSO\cite{picasso}) which all use
nuclear targets with an unpaired proton (in particular, they all
contain $^{19}F$).  As a consequence, this class of experiments have
been shown to rule out the scenario of Eq. (\ref{eq:spin_dependent})
also for $c^n\ll c^p$ when standard assumptions are made on the WIMP
local density and velocity distribution in our
Galaxy\cite{spin_gelmini}.

%%%%%%%%%%%%%%%%%%%%%%%%%%%%%%%%%%%%%%%%%%%%%%%%%%%%%%%
\begin{figure}
\begin{center}
\includegraphics[width=0.49\columnwidth, bb=73 193 513 636]{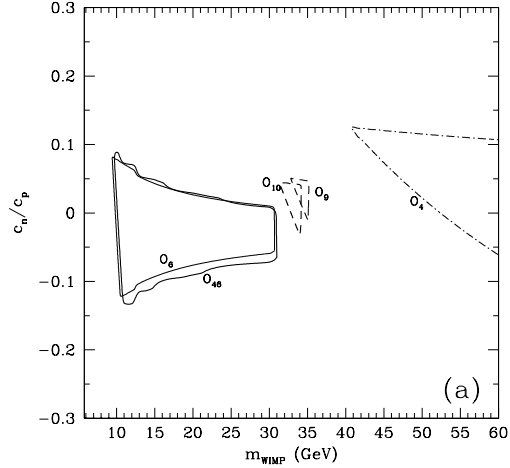}
\end{center}
\caption{Contour plot in the $m_{WIMP}$--$c^n_i/c^p_i$ plane for the
  compatibility factor ${\cal D}$ defined in Eqs. (5.1,5.2) of
  Ref.\cite{noi3}. ${\cal D}\le$1 implies compatibility between DAMA
  and other constraints.  The constant value ${\cal D}$=1 is shown for
  models ${\cal O}_i$, $i=6,46,9,10$ (which represent generalized
  spin--dependent interactions including explicit momentum and
  velocity dependence), while a value close to the minimum (${\cal
    D}$=1.7) is plotted for ${\cal O}_4$, which represents the
  standard spin--dependent interaction, for which DAMA and other
  constraints cannot be reconciled.}
\label{fig:generalized_spin}
\end{figure}
%%%%%%%%%%%%%%%%%%%%%%%%%%%%%%%%%%%%%%%%%%%%%%%%%%%%%%%

In light of this, in Ref. \cite{noi3} we extended the analysis of
spin--dependent WIMP--nucleus interactions. The most general
WIMP--nucleus spin--dependent interactions ca be singled out by making
use of the non--relativistic Effective Field Theory (EFT) approach of
Ref.\cite{haxton}. According to \cite{haxton} the most general
Hamiltonian density for the WIMP--nucleon process can be expressed in
terms of a combination of five Hermitian operators which act on the
two--particle Hilbert space spanned by tensor products of WIMP and
nucleon states; including terms that are at most linear in the nuclear
and WIMP spins and quadratic on the WIMP incoming velocity, the most
general Hamiltonian density describing the WIMP--nucleus interaction
can be written in terms of 15 non-relativistic quantum mechanical
operators. We used this approach to classify the most general
spin--dependent WIMP--nucleus interactions, and within this class of
models we discussed the viability of an interpretation of the DAMA
modulation result in terms of a WIMP signal, using a halo--independent
approach.

Our main conclusions were that, although several relativistic EFT's can
lead to a spin--dependent cross section, in some cases with an
explicit, non-negligible dependence on the WIMP incoming velocity,
three main scenarios can be singled out which approximately encompass
them all, and that only differ by their explicit dependence on the
transferred momentum. They are represented by models ${\cal O}_6$
($\simeq {\cal O}_{46}$), ${\cal O}_9$ ($\simeq {\cal O}_{10}$), and
${\cal O}_4$ in Fig. \ref{fig:generalized_spin}. For two of them
compatibility between DAMA and other constraints can be achieved for a
WIMP mass below 30 GeV, but only for a WIMP velocity distribution in
the halo of our Galaxy which departs from a Maxwellian. This is
achieved by combining a suppression of the WIMP effective coupling to
neutrons (to evade constraints from xenon and germanium detectors) to
an explicit quadratic or quartic dependence of the cross section on
the transferred momentum (that leads to a relative enhancement of the
expected rate off sodium in DAMA compared to that off fluorine in
droplet detectors and bubble chambers). For larger WIMP masses the
same scenarios are excluded by scatterings off iodine in COUPP.

The results summarized in this presentation show that, when different
new approaches such as the halo--dependence factorization,
non--relativistic EFT and inelastic scattering are combined together,
conclusions which are very different to what usually claimed in the
literature can be drawn. This shows that we are only starting now to
scratch the surface of the most general WIMP direct detection
parameter space.

\end{document}